%% file: setevolve.tex
\documentclass[sigconf]{acmart}
\include{commands}

\copyrightyear{2019} 
\acmYear{2019} 
\acmConference[CIKM'19]{The 28th ACM International Conference on Information and Knowledge Management}{November 3--7, 2019}{Beijing, China}
\acmPrice{15.00}
\acmDOI{10.1145/3357384.3358068}
\acmISBN{978-1-4503-6976-3/19/11}

\setcopyright{acmcopyright}
\newcommand{\astfootnote}[1]{
\let\oldthefootnote=\thefootnote
\setcounter{footnote}{1}
\renewcommand{\thefootnote}{\fnsymbol{footnote}}
\footnote{#1}
\let\thefootnote=\oldthefootnote
}

\begin{document}
\title[\textsc{SetEvolve}]{Query-Specific Knowledge Summarization with Entity Evolutionary Networks}
\author{Carl Yang\footnotemark, Lingrui Gan\setcounter{footnote}{0}\footnotemark, Zongyi Wang, Jiaming Shen, Jinfeng Xiao, Jiawei Han}
       \affiliation{
       \institution{University of Illinois, Urbana Champaign, 201 N Goodwin Ave, Urbana, IL 61801, USA}
       \institution{\{jiyang3, lgan6, zwang195, js2, jxiao13, hanj\}@illinois.edu}
       }

\setlength{\floatsep}{4pt plus 4pt minus 1pt}
\setlength{\textfloatsep}{4pt plus 2pt minus 2pt}
\setlength{\intextsep}{4pt plus 2pt minus 2pt}
\setlength{\dbltextfloatsep}{3pt plus 2pt minus 1pt}
\setlength{\dblfloatsep}{3pt plus 2pt minus 1pt}
\setlength{\abovecaptionskip}{3pt}
\setlength{\belowcaptionskip}{2pt}
\setlength{\abovedisplayskip}{2pt plus 1pt minus 1pt}
\setlength{\belowdisplayskip}{2pt plus 1pt minus 1pt}

\renewcommand\footnotetextcopyrightpermission[1]{}
\renewcommand{\shortauthors}{Carl Yang, Lingrui Gan et al.}
\newcommand{\independent}{\perp\mkern-9.5mu\perp}
\settopmatter{printacmref=true, printfolios=true}

\input{sec-abstract}
\maketitle
\renewcommand{\thefootnote}{\fnsymbol{footnote}}
\footnotetext[1]{Both authors contribute equally.}

\input{sec-intro}
\input{sec-model}

\input{sec-exp}

\input{sec-case}
\input{sec-con}

\bibliographystyle{ACM-Reference-Format}
\bibliography{setevolve}
\end{document}

%% file: commands.tex
\usepackage{booktabs} 
\usepackage{amsmath,url,graphicx,amssymb}
\usepackage{amsfonts}
\usepackage{ctable}
\usepackage{multirow}
\usepackage{algorithm}
\usepackage{algpseudocode}
\usepackage{pifont}
\usepackage{color}
\usepackage{subfigure}
\usepackage{enumitem}

\newcommand{\mylistbegin}{
  \begin{list}{$\bullet$}
   {
     \setlength{\itemsep}{-2pt}
     \setlength{\leftmargin}{1em}
     \setlength{\labelwidth}{1em}
     \setlength{\labelsep}{0.5em} } }
\newcommand{\mylistend}{
   \end{list}  }

\newcommand{\eg}{\textit{e.g.}}

\newcommand{\ie}{\textit{i.e.}}
\newcommand{\etc}{\textit{etc}}

%% file: sec-abstract.tex
\begin{abstract}
Given a query, unlike traditional IR that finds relevant documents or entities, in this work, we focus on retrieving both entities and their connections for insightful knowledge summarization. 
For example, given a query ``computer vision'' on a CS literature corpus, rather than returning a list of relevant entities like ``cnn'', ``imagenet'' and ``svm'', we are interested in the connections among them, and furthermore, the evolution patterns of such connections along particular ordinal dimensions such as time. 
Particularly, we hope to provide structural knowledge relevant to the query, such as ``svm'' is related to ``imagenet'' but not ``cnn''. Moreover, we aim to model the changing trends of the connections, such as ``cnn'' becomes highly related to ``imagenet'' after 2010, which enables the tracking of knowledge evolutions. 
In this work, to facilitate such a novel insightful search system, we propose \textsc{SetEvolve}, which is a unified framework based on nonparanomal graphical models for evolutionary network construction from large text corpora. 
Systematic experiments on synthetic data and insightful case studies on real-world corpora demonstrate the utility of \textsc{SetEvolve}.
\end{abstract}

\keywords{\normalsize knowledge summaries, network construction, evolution analysis}

%% file: sec-intro.tex
\section{Introduction}
\label{sec:intro}
Given a query, for knowledge summarization, it is often crucial to understand its relevant entities and their inherent connections. More interestingly, the entities and connections may evolve over time or other ordinal dimensions, indicating dynamic behaviors and changing trends. 
For example, in biological scientific literature, the study of a disease might focus on particular genes for a period of time, and then shift to some others due to a technology breakthrough. Capturing such entity connections and their changing trends can enable various tasks including the analysis of concept evolution, forecast of future events and detection of outliers.

\smallskip
\noindent \textbf{Related work.}
Traditional information retrieval (IR) aims at returning a ranked list of documents according to their relevance to the query \cite{baeza2011modern}. To understand the results and distill knowledge, users then need to pick out and read some of the documents, which requires tedious information processing and often leads to inaccurate conclusions. 
To deal with this, recent works on entity search \cite{jameel2017medmber, shen2018setsearch++} aim to search for entities instead of documents, but they only return lists of isolated entities, thus incapable of providing insights about entity connections. 
Existing works on graph keyword search \cite{kacholia2005bidirectional} and natural language processing \cite{falke2017bringing, Tan2016AceMapAN} have been using graph structures for query result or knowledge summarization, but they do not consider the evolution of entity connections.

\smallskip
\noindent \textbf{Present work.}
In this work, we advocate for the novel task of \textit{entity evolutionary network construction} for query-specific knowledge summarization, which aims to return a set of query-relevant entities together with their evolutionary connections modeled by a series of networks automatically constructed from large-scale text corpora. 
Mathematically, we model entities as variables in a complex dynamic system, and estimate the connections among them based on their discrete occurrence within the documents. 

Regarding techniques, recent existing works on network structure estimation have studied the inference of time-varying networks 
\cite{hallac2017network, tomasi2018latent}. However, in our novel problem setting, there are two unique challenges: 1) Identifying the query-relevant set of entities from text corpora and 2) Constructing the evolutionary entity connections based on discrete entity observations. 
To orderly deal with them, we develop \textsc{SetEvolve}, a unified framework based on the principled nonparanormal graphical models \cite{liu2009nonparanormal, gan2018bayesian}.

For the first challenge, we assume a satisfactory method has been given for the retrieval of a list of documents based on query relevance (\eg, BM25 \cite{baeza2011modern}) and develop a general post-hoc method based on document rank list cutoff for query-relevant entity set identification. The cutoff is efficiently computed towards entity-document support recovery for accurate network construction, with theoretical and empirical analysis on the existence of optimality. 

For the second challenge, we formulate the problem as an evolutionary graphical lasso, by modeling entities as variables in an evolving Markov random field, and detecting their inherent connections by estimating the underlying inverse covariance matrix. Moreover, we leverage the robust nonparanormal transformation to deal with the ordinal discrete entity observations in documents. Through theoretical analysis, we show that our model can capture the true conditional connections among entities, while its computational efficiency remains the same to standard graphical lasso. 

We evaluate \textsc{SetEvolve} on both synthetic networks and real-world datasets. 
On synthetic datasets of different sizes, evolution patterns and noisy observations., \textsc{SetEvolve} leads to significant improvements of 9\%-21\% on the standard F1 measure compared with the strongest baseline from the state-of-the-art.
Furthermore, on three real-world corpora, example evolutionary networks constructed by \textsc{SetEvolve} provide plausible summarizations of query-relevant knowledge that are rich, clear and readily interpretable. 

%% file: sec-model.tex
\section{SetEvolve}
\label{sec:model}

\subsection{Overview}
Given a query $\mathcal{Q}$ on a text corpus $\mathcal{C}$, we aim to provide a \textit{clear} and \textit{interpretable} summarization $\mathcal{S}$ over the retrieved knowledge $\mathcal{K}$, which is extracted from $\mathcal{C}$ based on the relevance to $\mathcal{Q}$. In principle, $\mathcal{S}$ should help users easily understand the key \textit{concepts} within $\mathcal{K}$, as well as their \textit{interactions} and \textit{evolutions}.

To achieve this goal, we get motivated by recent success on text summarization with concept maps \cite{falke2017bringing}, and propose to represent $\mathcal{S}$ as a series of concept networks $\mathcal{S}=\{\mathcal{N}_1, \ldots, \mathcal{N}_T\}$, by stressing the evolutionary nature of concept interactions.
In each network $\mathcal{N}_t \in \mathcal{S}$, $t$ denotes a window in an arbitrary ordinal dimension of interest, such as \textsf{time, product price, user age}, \etc. Without loss of generality, we will focus on \textsf{time} as the \textit{evolving dimension} in the following. We further decompose $\mathcal{N}_t$ into $\mathcal{N}_t= \{\mathcal{V}_t, \mathcal{E}_t\}$, where $\mathcal{V}_t$ is the set of relevant concepts, and $\mathcal{E}_t$ is the set of concept interactions, both in the time window denoted by $t$. 

In this work, we propose \textsc{SetEvolve}, a unified framework based on nonparanomal graphical models \cite{liu2009nonparanormal, 
falke2017bringing, hallac2017network, yang2017bi, gan2018bayesian}, for the novel problem of constructing query-specific knowledge summarization from free text corpora. 
In particular, empowered by recent available structured data of knowledge bases and well-developed techniques of entity discovery \cite{hasibi2017entity, Shen2017SetExpanCS}, we leverage \textit{entities} to represent concepts of interest, and derive a theoretically sound method to identify the stable set of query-relevant entities.
Moreover, to model concept interactions in a principled way, we consider the corresponding entities as interconnected variables in a complex system and leverage graphical models to infer their connection patterns based on their occurrence in the text corpus as variable observations. 
Finally, to capture knowledge evolution, we consider observations associated with time (or other evolving dimension of interest), and jointly estimate a series of networks for characterizing changing trends.

\subsection{Entity Set Identification}
To construct the knowledge summarization $\mathcal{S}$, we first need to identify a set of entities to represent the key concepts that users care about. 
Entity set search without the consideration of network evolution is well studied \cite{shen2018setsearch+}. 
Differently from them, in this work, we develop a general post-hoc entity set identification method based on the theoretical guarantees of nonparanomal graphical models for document-entity support recovery, as shown in Theorem \ref{thm:con}.
Particularly, we assume that based on a list of documents $\mathcal{D}=[d_1, d_2, \ldots]$ ranked by relevance, the \textit{optimal} query-relevant entity set $\mathcal{V}^*$ will appear consistently within the top ranked documents. 
In other words, as we bring in more documents, we get more complete entity sets and supportive information for constructing the entity network, but too many documents will bring in less relevant entities and redundant information.
Therefore, we aim to find an optimal cutoff on $\mathcal{D}$ to extract $\mathcal{V}^*$ and further facilitate accurate evolutionary network construction. 

Without loss of generality, we use the classic BM25 \cite{baeza2011modern} to retrieve documents $\mathcal{D}$, while various other advanced methods like \cite{wang2018lambdaloss} can be trivially plugged in to better meet users' particular information need regarding query relevance.
To extract entities from $\mathcal{D}$, we assume the availability of non-evolving knowledge bases and utilize an entity linking tool called CMNS \cite{hasibi2017entity} to convert $\mathcal{D}$ into an entity set list $\mathcal{V}_D=[\mathcal{V}_1, \mathcal{V}_2, \ldots]$, where $\mathcal{V}_i\in\mathcal{V}$ is the set of entities in $d_i\in\mathcal{D}$.
To quantify the optimality, we propose to track the following metric sequentially as documents are taken in one-by-one and use it to determine the optimal entity set $\mathcal{V}^*$:
\begin{equation}
\small
\gamma(n) = \frac{n}{\log|\mathcal{V}_1\cup\mathcal{V}_2\ldots\cup\mathcal{V}_n|}.
\end{equation}
where $|\mathcal{V}|$ is the size of an entity set $\mathcal{V}$, and $n$ is the number of documents taken in from the rank list $\mathcal{D}$.

The optimal set $\mathcal{V}^*$ extracted from $\mathcal{D}^*$ is determine based on two criteria: $1)$ the convergence of $\gamma'(n)$, which is used to measure the completeness of the entity set; and $2)$ $\gamma(n)$ is larger than a prefixed threshold (\eg, 10), which provides the theoretical guarantees in Theorem \ref{thm:con} for document-entity support recovery. 


\begin{figure}[h!]
\centering
\hspace{-8pt}
\subfigure[PubMed]{
\includegraphics[width=0.23\textwidth]{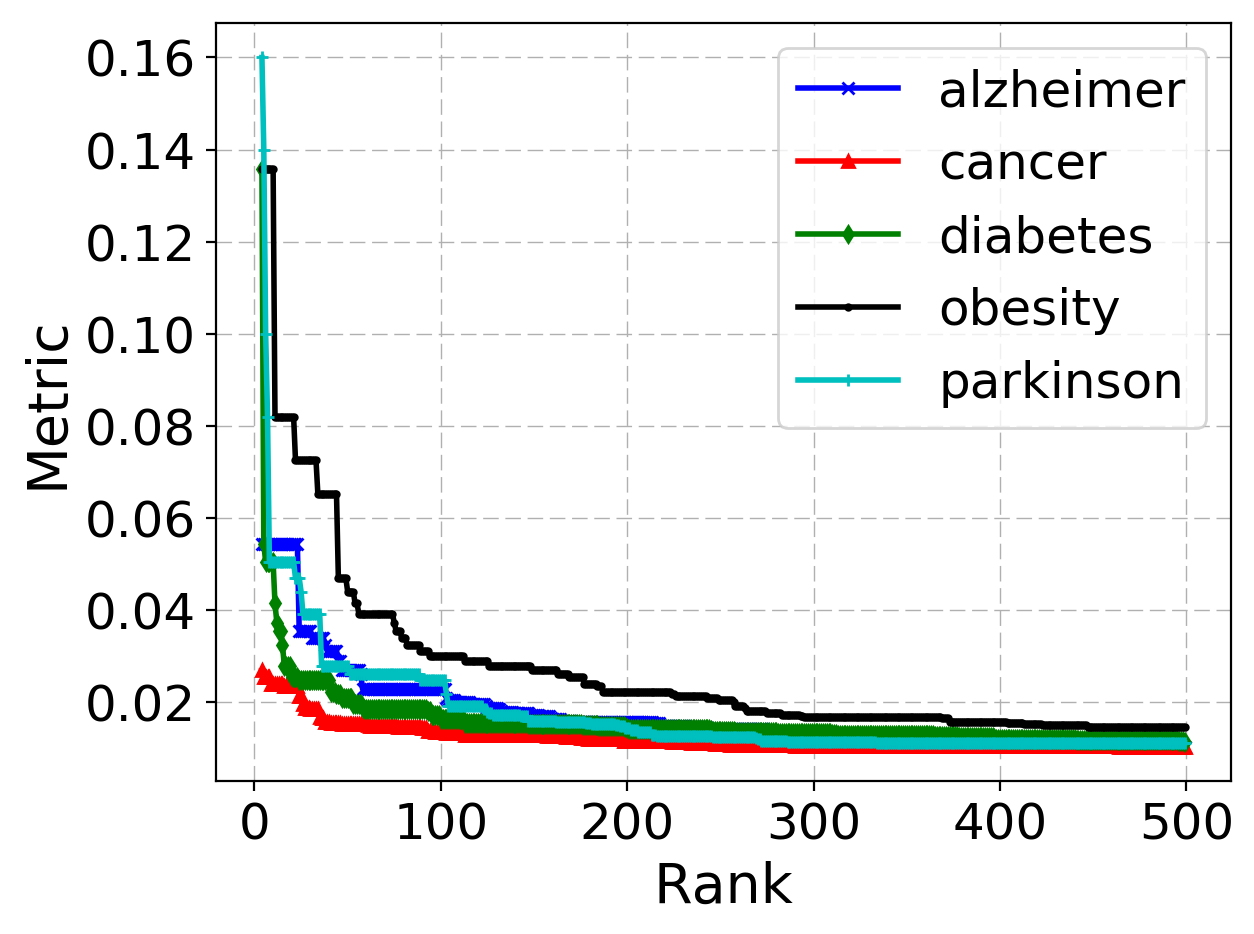}}
\hspace{-8pt}
\subfigure[Yelp]{
\includegraphics[width=0.23\textwidth]{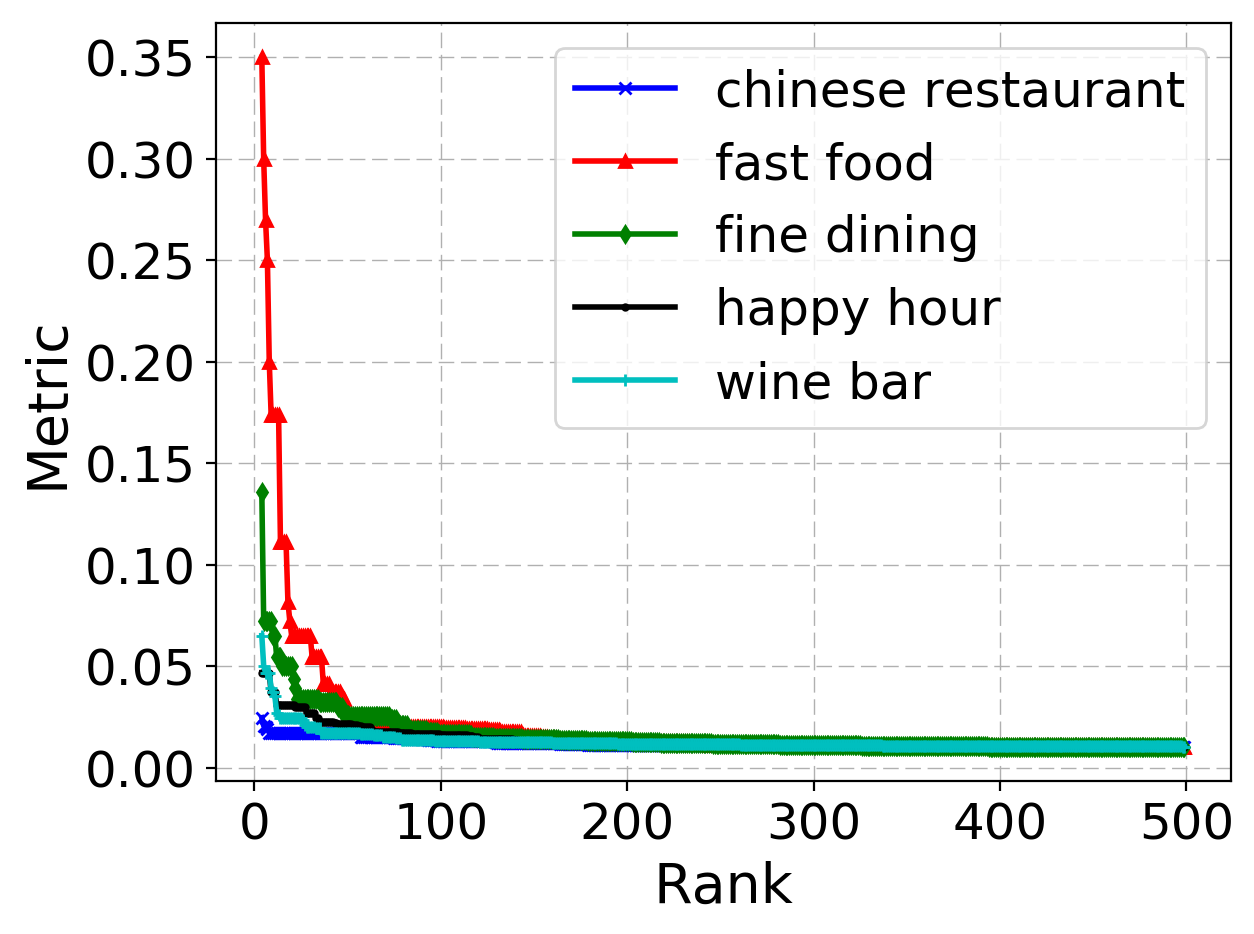}}
\hspace{-8pt}
\caption{\textbf{Entity-document support consistency (holder).}}
\label{fig:curve}
\end{figure}

Figure \ref{fig:curve} shows $\gamma'(n)$ computed on the PubMed and Yelp datasets with different queries. As we can see, the derivation of $\gamma(n)$ often converges to 0 rapidly, which corroborates our assumption of entity consistency. Particularly, as we take in more documents, the total number of entities converges, so we can cutoff the document list at $\mathcal{D}^*$ once the aforementioned two criteria are met, and use $\mathcal{D}^*$ to identify $\mathcal{V}^*$ and construct the evolutionary networks over $\mathcal{V}^*$.

\subsection{Evolutionary Network Construction}
After fixing $\mathcal{V}^*$, we consider the retrieval of connections among $\mathcal{V}^*$. 
A straightforward way to construct links in an entity network is to compute and threshold the document-level entity co-occurrence as suggested in \cite{peng2018large}. However, it is not clear how to weigh the links and set the thresholds. More importantly, because the thresholding method \cite{peng2018large} only models the pairwise \emph{marginal} dependence between entities $v_i$ and $v_j$, and does not consider the interactions between $\{v_i, v_j\}$ and all the other entities $v_{-\{i,j\}}$, the links generated may be messy and less insightful, as shown in \cite{shen2018setsearch+}.
 
To find the essential entity connections in a principled way, we propose to uncover the underlying connection patterns among entities as a graphical model selection problem \cite{friedman2008sparse, gan2018bayesian}. 
Assume we have $n$ observations $V^1, ..., V^n$ of a set of entities $\mathcal{V}=\{v_1, \ldots, v_m\}$. In a standard GGM, we assume the $n$ observations identically and independent distributed (\textit{iid}) follow a multivariate Gaussian distribution $V^1, \ldots, V^n \sim \mathcal{N}(0, \Theta^{-1})$, and our target is to detect the support of the precision matrix $\Theta$. It is well-known that $\Theta_{ij} = 0$ in standard GGM is equivalent to the conditional independence of $v_i$ and $v_j$, given all other variables, \ie, $v_i \not\!\perp\!\!\!\perp v_j\mid v_{-\{i,j\}}$. If the Gaussianity and \textit{iid} assumptions of GGM hold, insightful conditional dependence network links can be generated systematically.

However, the GGM assumptions do not exactly fit the context we consider, \ie, entity evolutionary network construction, because 1) GGM assumes observations are Gaussian, but entity occurrences in documents are discrete; 2) The entity network is evolving over time, so the \textit{iid} assumption does not hold.

Motivated by previous works in graphical models that address the non-Gaussian data \cite{liu2009nonparanormal} and time dependent data \cite{zhou2010time, kolar2011time}, we propose an \emph{evolutionary nonparanormal graphical model}, which detects the conditional dependence structure even when data is both \emph{discrete} and \emph{evolving}. Without loss of generality, in the following, we use \textit{time} as the evolving dimension to describe our model.

Denote the $m$-dimensional discrete observation at time $t$ as $Y^{(t)} = (Y^{(t)}_1, ..., Y^{(t)}_m)$. In the proposed model, we assume $Y^{(t)}$ satisfies the following relationship:
\begin{equation}
\small
f(Y^{(t)}) \sim \mathcal{N}(0, \Theta(t)^{-1}), 
\label{eq:model}
\end{equation}
where $f(\cdot)$ is an $m$-dimensional Gaussian copula transformation function defined in Eq.~(\ref{eq:copula})  and $\Theta(t)$ is assumed to evolve smoothly from $t$ to $t+1$. A definition of the smoothness assumption of the evolution pattern we use is as Assumption S in \cite{kolar2011time}, where the smoothness is quantified through the boundedness of the first and second derivative of the changes in $\Theta(t)_{i,j}$ over time $t$. No further assumption on the parameter (distribution) form of the evolution pattern is required for the model.

As shown in the following proposition, the model \eqref{eq:model} can capture the conditional dependence links even when data is \emph{evolving} and \emph{discrete}. The proposition can be shown through standard matrix calculation and we refer to Lemma 2 in \cite{liu2009nonparanormal} for the proof. 
\vspace{-0.1cm}
\begin{proposition}
If $\Theta(t)_{ij} = 0$, $Y^{(t)}_i$ and $Y^{(t)}_j$ are conditionally independent, i.e., $Y^{(t)}_i \independent Y^{(t)}_j\mid Y^{(t)}_{- \{i,j\}}$.
\label{prop:ind}
\end{proposition}
\vspace{-0.1cm}


Motivated by \cite{liu2009nonparanormal}, we utilize a Gaussian copula transformation function $f(\cdot)$ to handle the non-Gaussianity in the data, and the function $f(\cdot)$ is defined as follows:
\begin{equation}
\small
f_j(x) = u_j + \sigma_j  \Phi^{-1}(F_j(x)).
\label{eq:copula}
\end{equation}
In \eqref{eq:copula}, $u_j$ and $\sigma_j$ are the empirical mean and empirical standard deviation of $Y_j$ respectively, $\Phi^{-1}(\cdot)$ is the quantile function of the standard norm distribution, and $F_j(x)$ is the Winsorized estimator of the empirical distribution suggested in \cite{liu2009nonparanormal} and defined as:
\begin{equation}
\small
F_j(X) = \begin{cases}
\delta_n, \quad \text{ if } \hat{F}_j(X) < \delta_n; \\
\hat{F}_j(X), \quad \text{ if }  \delta_n \leq \hat{F}_j(X) \leq 1-\delta_n; \\
1-\delta_n, \quad \text{ if } \hat{F}_j(X) > 1-\delta_n,
\end{cases}
\end{equation}
where $\delta_n = \frac{1}{4n^{1/4}\sqrt{\pi\log n}}$ and $\hat{F}_j(X)$ is the empirical cumulative distribution function. 

This transformation allows us to estimate the evolving pattern of $\Theta(t)$ through a kernel based method \cite{zhou2010time,kolar2011time}, where our target is to minimize the following kernel-based objective function:
\begin{equation}
\small
\hat{\Theta}(t) = \arg\min_{\Theta(t) \succ 0}\left\{-\log |\Theta| + \text{tr}(\Theta S_f(t)) + \lambda|\Theta|\right\}.
\label{eq:est}
\end{equation}
$S_f(t) = \sum_{t'}w^t_{t'} f(Y^{(t')})f(Y^{(t')})'$ is a kernel estimator of the sample covariance matrix at time $t$, with weights $w^t_{t'}$ defined as:
\begin{equation}
\small
w^t_{t'} = \frac{K_h(|t - t'|)}{\sum_{t'} K_h(|t - t'|)},
\label{eq:weight}
\end{equation} 
where $K_h(\cdot)$ is a kernel function such as the box kernel $K_h(x) = \frac{1}{2}\times \mathbb{I}\{\frac{x}{h} \in [-1, 1]\}$. 


\subsection{Theoretical Analysis}
In the following theorem, we show the accuracy of \textsc{SetEvolve} in detecting the true entity links. 
\vspace{-0.1cm}
\begin{theorem}
With the same assumptions of \cite{kolar2011time} on the true evolving graphs, define the maximum graph node degree as $d$. Suppose $S_f(t)$ is estimated using a kernel with bandwidth $h = O(n^{-1/6})$. If the number of documents $n$ satisfies $n > C(d^2\log m)^3$ for a sufficiently large constant $C$ and we choose tuning parameter $\lambda = O(n^{-1/6}\log n\sqrt{\log m})$, then our estimation procedure can detect the links correctly with probability converging to $1$. 
\label{thm:con}
\end{theorem}

\vspace{-0.1cm}
\begin{proof}
$||S_f(t) - (\Theta^0(t))^{-1}||_\infty \leq ||S_f(t) - ES_f(t)||_\infty  + ||ES_f(t) - (\Theta^0(t))^{-1}||_\infty$, where the first term is $O_p(\sqrt{\frac{\log m\log^2n}{hn^{1/2}}})$ by Theorem 4 in \cite{liu2009nonparanormal} and the second is $O_p(h)$ by Lemma 11 in \cite{kolar2011time}. If the bandwidth $h$ is chosen as $h = O(n^{-1/6})$, we have $||S_f(t) - (\Theta^0(t))^{-1}||_\infty = O_p(\sqrt{\frac{\log m\log^2n}{n^{1/3}}})$, \ie, $||S_f(t) - (\Theta^0(t))^{-1}||_\infty <\sqrt{\frac{\log m\log^2n}{n^{1/3}}}$. 

Assuming conditions in Theorem \ref{thm:con} hold, it implies Proposition 3 and 4 in \cite{kolar2011time} hold with $w(n,m) = n^{-1/6}\log n\sqrt{\log m}$ and $\lambda = Cn^{-1/6}\log n\sqrt{\log m}$. We then apply Proposition 2 in \cite{kolar2011time} to conclude our proof on link detection accuracy.
\end{proof}
\vspace{-0.1cm}

In terms of the computation efficiency, in \textsc{SetEvolve}, we first compute $S_f(t)$ using equation \eqref{eq:weight}, and then compute $\Theta(t)$ for each $t$ using the state-of-the-art Graphical Lasso algorithm \cite{friedman2008sparse}. 
The computation complexity is thus exactly same as the state-of-the-art evolutionary network inference algorithms \cite{hallac2017network, tomasi2018latent}.

%

%% file: sec-exp.tex
\section{Synthetic Experiments}
Since there is no ground-truth for knowledge summarization with evolutionary networks, we follow the common practice in recent works on network inference \cite{hallac2017network, tomasi2018latent} to demonstrate the effectiveness of \textsc{SetEvolve} through comprehensive synthetic experiments.


\smallskip
\noindent \textbf{Data Generation.}
Following \citet{hallac2017network}, we generate two evolutionary networks of sizes $20$ and $100$ and then randomly generate the ground-truth covariance and sample discrete observations over 100 timestamps, where the global and local evolutions occur at time $t=50$. At each $t$, we generate 10 independent samples from the true distribution, with maximum variable value set to 10. 

%

\smallskip
\noindent \textbf{Compared Algorithms.}
We compare \textsc{SetEvolve} to two baselines: 1) \textit{Static} \cite{liu2009nonparanormal}, which applies the nonparanormal transformation for discrete observations, but aggregate observations from all timestamps to infer a single static network without evolutions, and 2) \textit{TVGL} \cite{hallac2017network}, which leverages graphical lasso for evolutionary network construction but does not consider discrete observations.


\smallskip
\noindent \textbf{Performance Evaluations.}
Table \ref{tab:perform} shows the performance of compared methods.
Following \cite{hallac2017network, tomasi2018latent}, we compute macro-F1 scores for link reconstruction over evolutionary networks. 
As we can see, all algorithms perform better on smaller networks, partly because the observations are denser and the essential correlations are easier to capture. 
Moreover, all algorithms perform better with local evolutions, because the overall changes are smaller, while \textit{TVGL} and \textsc{SetEvolve} perform better on global evolutions compared with \textit{Static}. 
Finally, \textsc{SetEvolve} consistently outperforms both \textit{Static} and \textit{TVGL} on networks of different sizes and evolution types, which indicates its effectiveness in detecting correlations among evolutionary discrete variables. 
The scores of TVGL are lower than those in the original work due to the discreteness of variables. 
Besides accuracy, \textsc{SetEvolve} can construct networks with fewest links, generating clear views of networks for deriving the most essential entity connections. 

\begin{table}[h]
 \centering

    \scalebox{0.85}{
 \begin{tabular}{|c|cc|cc|cc|}
 \hline
\multirow{2}{*}{Networks}&\multicolumn{2}{c|}{Static}&\multicolumn{2}{c|}{TVGL}&\multicolumn{2}{c|}{SetEvolve}\\
\cline{2-7}
&F1&$|\mathcal{E}|$&F1&$|\mathcal{E}|$&F1&$|\mathcal{E}|$\\
\hline
20-global & 0.3698 & 97 & 0.4058 & 104 & \textbf{0.4592} & \textbf{58} \\
20-local & 0.4567 & 90 & 0.5148 & 85 & \textbf{0.6260} & \textbf{35} \\
\hline
100-global & 0.3239 & 1573 & 0.3372 & 2020 & \textbf{0.3670} & \textbf{687} \\
100-local & 0.3684 & 1548 & 0.3683 & 1895 & \textbf{0.4045} & \textbf{637} \\
\hline
 \end{tabular}
 }
 \caption{\label{tab:perform}\textbf{Link detection performance.}}
 \vspace{-5pt}
\end{table}

\begin{figure}[h!]
\centering
\vspace{-5pt}
\hspace{-10pt}
\subfigure{
\includegraphics[width=0.21\textwidth]{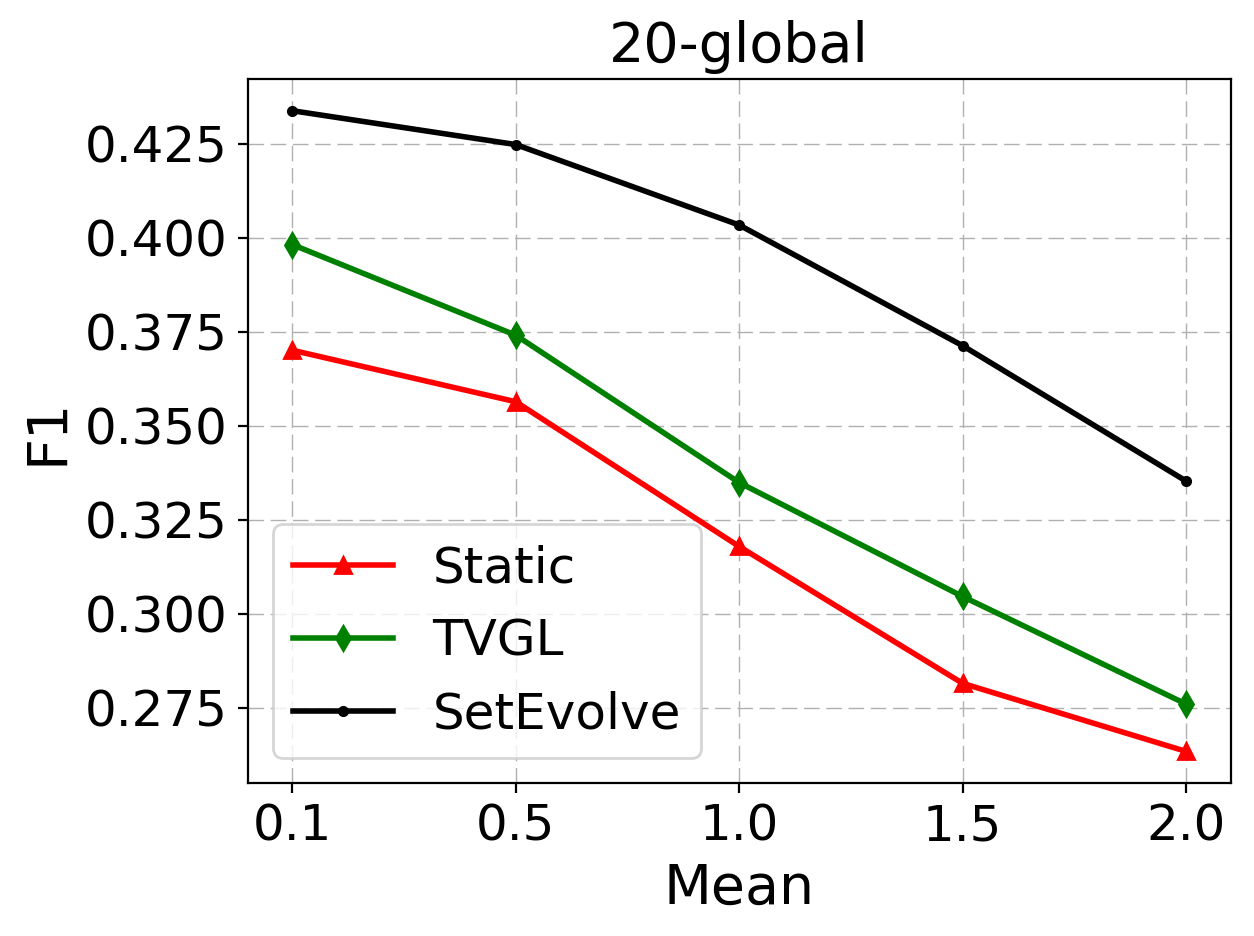}\vspace{-10pt}}
\subfigure{
\includegraphics[width=0.21\textwidth]{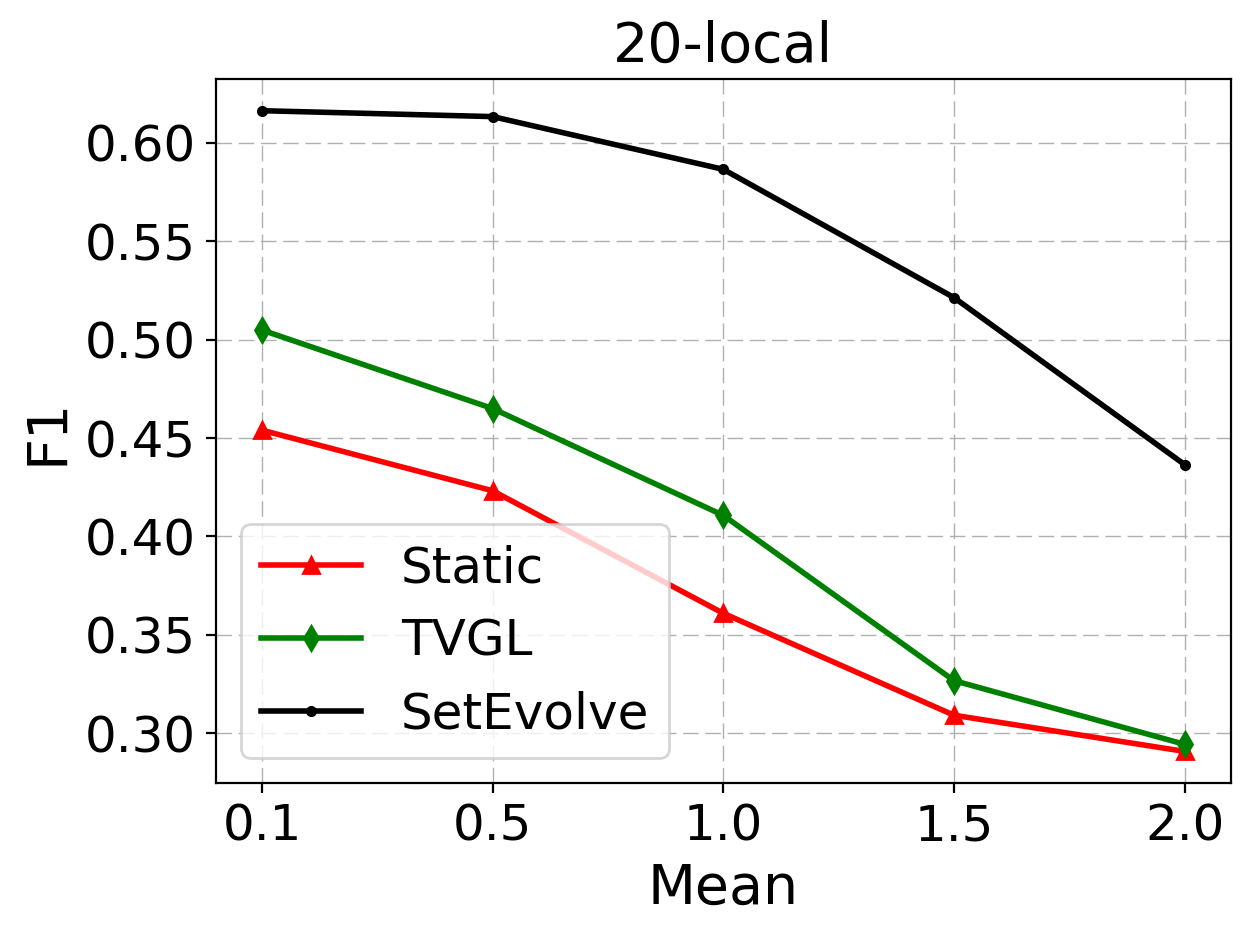}\vspace{-10pt}}
\vspace{-5pt}
\caption{\textbf{Performance with noisy observations.}}
\vspace{-10pt}
\label{fig:noise}
\end{figure}

We evaluate the robustness of \textsc{SetEvolve} by randomly adding Poisson noises to synthetic observations and show results in Figure \ref{fig:noise}.
We also find that runtimes of \textit{TVGL} and \textsc{SetEvolve} are roughly the same, which indicates the good scalability of \textsc{SetEvolve}. 

%% file: sec-case.tex
\section{Case Studies}
We show the practical utility of \textsc{SetEvolve} on three datasets: \textit{PubMed} with $14$ GB texts of $3.3M$ articles\footnote{\scriptsize https://www.ncbi.nlm.nih.gov/pubmed/}, \textit{Yelp} with $5.4$ GB texts of $6M$ reviews\footnote{\scriptsize https://www.yelp.com/dataset/challenge}, and \textit{Semantic} with $19$ GB texts of $20M$ papers\footnote{\scriptsize https://labs.semanticscholar.org/corpus/corpus/archive}. 

\begin{figure}[h!]
\centering
\vspace{-5pt}
\hspace{-15pt}
\subfigure[Year: 1985]{
\includegraphics[width=0.15\textwidth]{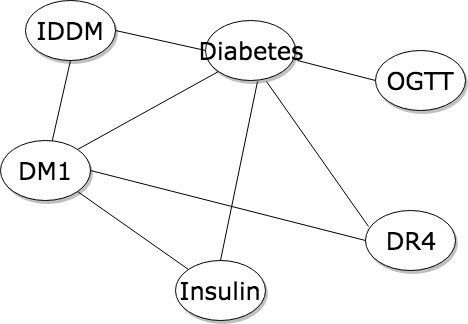}}
\subfigure[Year: 2000]{
\includegraphics[width=0.15\textwidth]{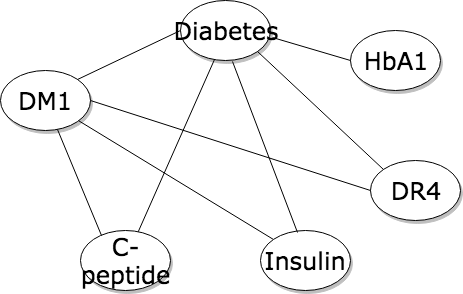}}
\subfigure[Year: 2015]{
\includegraphics[width=0.15\textwidth]{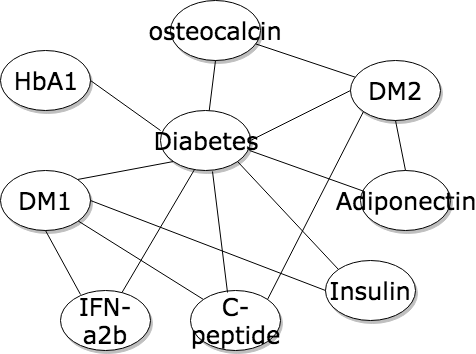}}
\hspace{-10pt}
\vspace{-5pt}
\caption{\textbf{Querying \textit{Diabetes} on \textit{PubMed}.}}
\vspace{-5pt}
\label{fig:pubmed}
\end{figure}

\begin{figure}[h!]
\centering
\hspace{-15pt}
\subfigure[Rating: 1]{
\includegraphics[width=0.15\textwidth]{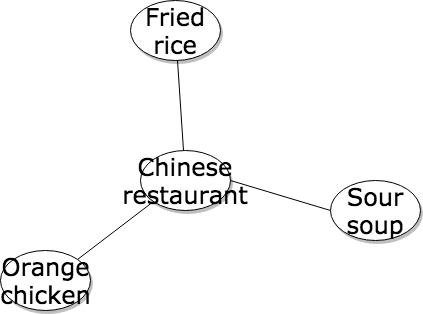}}
\subfigure[Rating: 3]{
\includegraphics[width=0.15\textwidth]{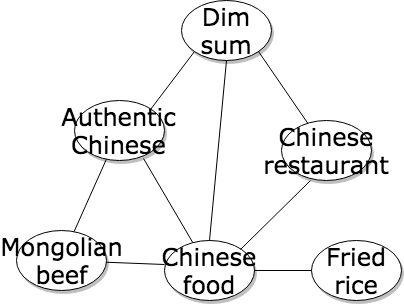}}
\subfigure[Rating: 5]{
\includegraphics[width=0.15\textwidth]{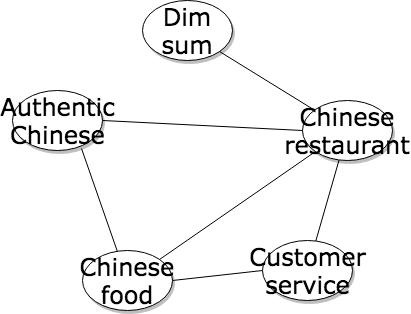}}
\hspace{-10pt}
\vspace{-5pt}
\caption{\textbf{Querying \textit{Chinese food} on \textit{Yelp}.}}
\vspace{-5pt}
\label{fig:yelp}
\end{figure}

\begin{figure}[h!]
\centering
\hspace{-15pt}
\subfigure[Citation: 10]{
\includegraphics[width=0.15\textwidth]{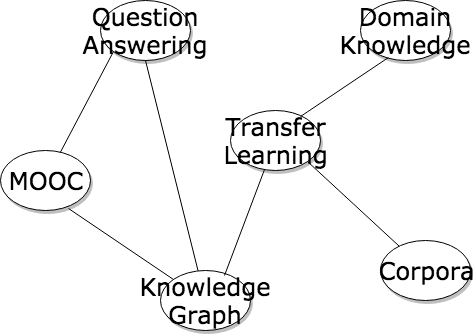}}
\subfigure[Citation: 50]{
\includegraphics[width=0.15\textwidth]{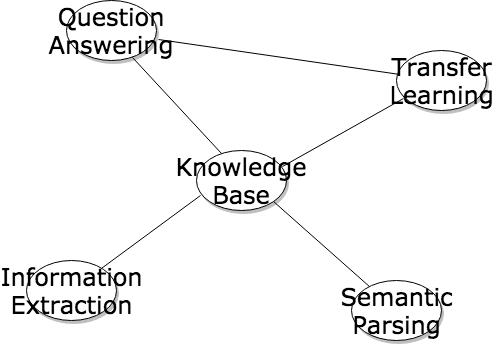}}
\subfigure[Citation: 100]{
\includegraphics[width=0.15\textwidth]{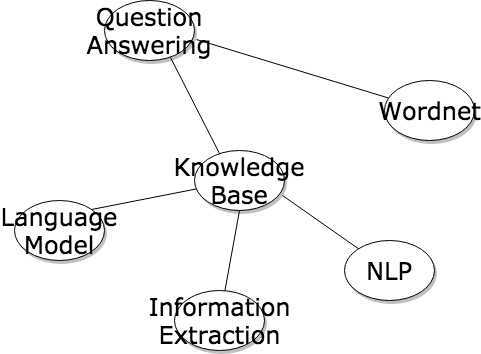}}
\hspace{-10pt}
\vspace{-5pt}
\caption{\textbf{Querying \textit{Knowledge summarization} on \textit{Semantic}.}}
\label{fig:semantic}
\end{figure}

Figure \ref{fig:pubmed}, \ref{fig:yelp} and \ref{fig:semantic} show the constructed networks on each corpus. 
The results are insightful.
Figure \ref{fig:pubmed}, for example, shows that the long-used \textsf{OGTT} test for \textsf{Diabetes} was replaced by a new standard \textsf{HbA1} test in the 2000's, and \textsf{Type 2 Diabetes (DM2)} draws more research attention with newly found proteins like \textsf{Osteocalcin} in recent years. 
In Figure \ref{fig:yelp}, we notice that \textsf{Chinese restaurants} get poorly rated often due to the quality of classic dishes (\eg, \textsf{orange chicken}, \textsf{sour soup}), while the higher ratings focus on \textsf{authentication} and \textsf{customer service}.
In Figure \ref{fig:semantic}, we observe that papers about \textsf{MOOC} and \textsf{transfer learning} have accumulated fewer citations, partly because these fields are quite new, whereas papers about \textsf{question answering}, \textsf{knowledge base} and \textsf{language model} are usually cited more.
We will release the full implementation of our framework and interested readers can pose queries to find more insightful knowledge.

%% file: sec-con.tex
\vspace{-0.1cm}
\section{Conclusion}
We propose \textsc{SetEvolve}, a unified framework for query-specific knowledge summarization from large text corpora. Built on the principles of nonparanormal graphical models, \textsc{SetEvolve} identifies query-relevant entity sets and constructs evolutionary entity networks with theoretical guarantees. Its effectiveness is corroborated with rich synthetic experiments and insightful case studies.

%% file: setevolve.bbl

\begin{thebibliography}{00}


\ifx \showCODEN    \undefined \def \showCODEN     #1{\unskip}     \fi
\ifx \showDOI      \undefined \def \showDOI       #1{{\tt DOI:}\penalty0{#1}\ }
  \fi
\ifx \showISBNx    \undefined \def \showISBNx     #1{\unskip}     \fi
\ifx \showISBNxiii \undefined \def \showISBNxiii  #1{\unskip}     \fi
\ifx \showISSN     \undefined \def \showISSN      #1{\unskip}     \fi
\ifx \showLCCN     \undefined \def \showLCCN      #1{\unskip}     \fi
\ifx \shownote     \undefined \def \shownote      #1{#1}          \fi
\ifx \showarticletitle \undefined \def \showarticletitle #1{#1}   \fi
\ifx \showURL      \undefined \def \showURL       #1{#1}          \fi
\providecommand\bibfield[2]{#2}
\providecommand\bibinfo[2]{#2}
\providecommand\natexlab[1]{#1}
\providecommand\showeprint[2][]{arXiv:#2}

\bibitem[\protect\citeauthoryear{Baeza-Yates, Ribeiro,
  et~al\mbox{.}}{Baeza-Yates et~al\mbox{.}}{2011}]%
        {baeza2011modern}
\bibfield{author}{\bibinfo{person}{Ricardo Baeza-Yates},
  \bibinfo{person}{Berthier de Ara{\'u}jo~Neto Ribeiro}, {and}
  \bibinfo{person}{others}.} \bibinfo{year}{2011}\natexlab{}.
\newblock \bibinfo{booktitle}{{\em Modern information retrieval}}.
\newblock \bibinfo{publisher}{New York: ACM Press; Harlow, England:
  Addison-Wesley,}.
\newblock


\bibitem[\protect\citeauthoryear{Falke and Gurevych}{Falke and
  Gurevych}{2017}]%
        {falke2017bringing}
\bibfield{author}{\bibinfo{person}{Tobias Falke} {and} \bibinfo{person}{Iryna
  Gurevych}.} \bibinfo{year}{2017}\natexlab{}.
\newblock \showarticletitle{Bringing Structure into Summaries: Crowdsourcing a
  Benchmark Corpus of Concept Maps}. In \bibinfo{booktitle}{{\em ACL}}.
\newblock


\bibitem[\protect\citeauthoryear{Friedman, Hastie, and Tibshirani}{Friedman
  et~al\mbox{.}}{2008}]%
        {friedman2008sparse}
\bibfield{author}{\bibinfo{person}{Jerome Friedman}, \bibinfo{person}{Trevor
  Hastie}, {and} \bibinfo{person}{Robert Tibshirani}.}
  \bibinfo{year}{2008}\natexlab{}.
\newblock \showarticletitle{Sparse inverse covariance estimation with the
  graphical lasso}. In \bibinfo{booktitle}{{\em Biostatistics}}.
\newblock


\bibitem[\protect\citeauthoryear{Gan, Narisetty, and Liang}{Gan
  et~al\mbox{.}}{2018}]%
        {gan2018bayesian}
\bibfield{author}{\bibinfo{person}{Lingrui Gan}, \bibinfo{person}{Naveen~N
  Narisetty}, {and} \bibinfo{person}{Feng Liang}.}
  \bibinfo{year}{2018}\natexlab{}.
\newblock \showarticletitle{Bayesian Regularization for Graphical Models with
  Unequal Shrinkage}. In \bibinfo{booktitle}{{\em JASA}}.
\newblock


\bibitem[\protect\citeauthoryear{Hallac, Park, Boyd, and Leskovec}{Hallac
  et~al\mbox{.}}{2017}]%
        {hallac2017network}
\bibfield{author}{\bibinfo{person}{David Hallac}, \bibinfo{person}{Youngsuk
  Park}, \bibinfo{person}{Stephen Boyd}, {and} \bibinfo{person}{Jure
  Leskovec}.} \bibinfo{year}{2017}\natexlab{}.
\newblock \showarticletitle{Network Inference via the Time-Varying Graphical
  Lasso}. In \bibinfo{booktitle}{{\em KDD}}.
\newblock


\bibitem[\protect\citeauthoryear{Hasibi, Balog, and Bratsberg}{Hasibi
  et~al\mbox{.}}{2017}]%
        {hasibi2017entity}
\bibfield{author}{\bibinfo{person}{Faegheh Hasibi}, \bibinfo{person}{Krisztian
  Balog}, {and} \bibinfo{person}{Svein~Erik Bratsberg}.}
  \bibinfo{year}{2017}\natexlab{}.
\newblock \showarticletitle{Entity linking in queries: Efficiency vs.
  effectiveness}. In \bibinfo{booktitle}{{\em ECIR}}.
\newblock


\bibitem[\protect\citeauthoryear{Jameel, Bouraoui, and Schockaert}{Jameel
  et~al\mbox{.}}{2017}]%
        {jameel2017medmber}
\bibfield{author}{\bibinfo{person}{Shoaib Jameel}, \bibinfo{person}{Zied
  Bouraoui}, {and} \bibinfo{person}{Steven Schockaert}.}
  \bibinfo{year}{2017}\natexlab{}.
\newblock \showarticletitle{MEmbER: Max-Margin Based Embeddings for Entity
  Retrieval}. In \bibinfo{booktitle}{{\em SIGIR}}.
\newblock


\bibitem[\protect\citeauthoryear{Kacholia, Pandit, Chakrabarti, Sudarshan,
  Desai, and Karambelkar}{Kacholia et~al\mbox{.}}{2005}]%
        {kacholia2005bidirectional}
\bibfield{author}{\bibinfo{person}{Varun Kacholia}, \bibinfo{person}{Shashank
  Pandit}, \bibinfo{person}{Soumen Chakrabarti}, \bibinfo{person}{S Sudarshan},
  \bibinfo{person}{Rushi Desai}, {and} \bibinfo{person}{Hrishikesh
  Karambelkar}.} \bibinfo{year}{2005}\natexlab{}.
\newblock \showarticletitle{Bidirectional expansion for keyword search on graph
  databases}. In \bibinfo{booktitle}{{\em VLDB}}.
\newblock


\bibitem[\protect\citeauthoryear{Kolar and Xing}{Kolar and Xing}{2011}]%
        {kolar2011time}
\bibfield{author}{\bibinfo{person}{Mladen Kolar} {and} \bibinfo{person}{Eric
  Xing}.} \bibinfo{year}{2011}\natexlab{}.
\newblock \showarticletitle{On time varying undirected graphs}. In
  \bibinfo{booktitle}{{\em ICAIS}}.
\newblock


\bibitem[\protect\citeauthoryear{Liu, Lafferty, and Wasserman}{Liu
  et~al\mbox{.}}{2009}]%
        {liu2009nonparanormal}
\bibfield{author}{\bibinfo{person}{Han Liu}, \bibinfo{person}{John Lafferty},
  {and} \bibinfo{person}{Larry Wasserman}.} \bibinfo{year}{2009}\natexlab{}.
\newblock \showarticletitle{The nonparanormal: Semiparametric estimation of
  high dimensional undirected graphs}. In \bibinfo{booktitle}{{\em JMLR}}.
\newblock


\bibitem[\protect\citeauthoryear{Peng, Li, He, Liu, Bao, Wang, Song, and
  Yang}{Peng et~al\mbox{.}}{2018}]%
        {peng2018large}
\bibfield{author}{\bibinfo{person}{Hao Peng}, \bibinfo{person}{Jianxin Li},
  \bibinfo{person}{Yu He}, \bibinfo{person}{Yaopeng Liu},
  \bibinfo{person}{Mengjiao Bao}, \bibinfo{person}{Lihong Wang},
  \bibinfo{person}{Yangqiu Song}, {and} \bibinfo{person}{Qiang Yang}.}
  \bibinfo{year}{2018}\natexlab{}.
\newblock \showarticletitle{Large-Scale Hierarchical Text Classification with
  Recursively Regularized Deep Graph-CNN}. In \bibinfo{booktitle}{{\em WWW}}.
\newblock


\bibitem[\protect\citeauthoryear{Shen, Wu, Lei, Shang, Ren, and Han}{Shen
  et~al\mbox{.}}{2017}]%
        {Shen2017SetExpanCS}
\bibfield{author}{\bibinfo{person}{Jiaming Shen}, \bibinfo{person}{Zeqiu Wu},
  \bibinfo{person}{Dongming Lei}, \bibinfo{person}{Jingbo Shang},
  \bibinfo{person}{Xiang Ren}, {and} \bibinfo{person}{Jiawei Han}.}
  \bibinfo{year}{2017}\natexlab{}.
\newblock \showarticletitle{SetExpan: Corpus-Based Set Expansion via Context
  Feature Selection and Rank Ensemble}. In \bibinfo{booktitle}{{\em
  ECML/PKDD}}.
\newblock


\bibitem[\protect\citeauthoryear{Shen, Xiao, He, Shang, Sinha, and Han}{Shen
  et~al\mbox{.}}{2018a}]%
        {shen2018setsearch+}
\bibfield{author}{\bibinfo{person}{Jiaming Shen}, \bibinfo{person}{Jinfeng
  Xiao}, \bibinfo{person}{Xinwei He}, \bibinfo{person}{Jingbo Shang},
  \bibinfo{person}{Saurabh Sinha}, {and} \bibinfo{person}{Jiawei Han}.}
  \bibinfo{year}{2018}\natexlab{a}.
\newblock \showarticletitle{Entity set search of scientific literature}. In
  \bibinfo{booktitle}{{\em SIGIR}}.
\newblock


\bibitem[\protect\citeauthoryear{Shen, Xiao, Zhang, Yang, Shang, Sinha, Ping,
  Lu, and Han}{Shen et~al\mbox{.}}{2018b}]%
        {shen2018setsearch++}
\bibfield{author}{\bibinfo{person}{Jiaming Shen}, \bibinfo{person}{Jinfeng
  Xiao}, \bibinfo{person}{Yu Zhang}, \bibinfo{person}{Carl Yang},
  \bibinfo{person}{Jingbo Shang}, \bibinfo{person}{Saurabh Sinha},
  \bibinfo{person}{Peipei Ping}, \bibinfo{person}{Zhiyong Lu}, {and}
  \bibinfo{person}{Jiawei Han}.} \bibinfo{year}{2018}\natexlab{b}.
\newblock \showarticletitle{SetSearch+: Entity-Set-Aware Search and Mining for
  Scientific Literature}. In \bibinfo{booktitle}{{\em KDD}}.
\newblock


\bibitem[\protect\citeauthoryear{Tan, Liu, Mao, Guo, Shen, and Wang}{Tan
  et~al\mbox{.}}{2016}]%
        {Tan2016AceMapAN}
\bibfield{author}{\bibinfo{person}{Zhaowei Tan}, \bibinfo{person}{Changfeng
  Liu}, \bibinfo{person}{Yuning Mao}, \bibinfo{person}{Yunqi Guo},
  \bibinfo{person}{Jiaming Shen}, {and} \bibinfo{person}{Xinbing Wang}.}
  \bibinfo{year}{2016}\natexlab{}.
\newblock \showarticletitle{AceMap: A Novel Approach towards Displaying
  Relationship among Academic Literatures}. In \bibinfo{booktitle}{{\em WWW}}.
\newblock


\bibitem[\protect\citeauthoryear{Tomasi, Tozzo, Salzo, and Verri}{Tomasi
  et~al\mbox{.}}{2018}]%
        {tomasi2018latent}
\bibfield{author}{\bibinfo{person}{Federico Tomasi}, \bibinfo{person}{Veronica
  Tozzo}, \bibinfo{person}{Saverio Salzo}, {and} \bibinfo{person}{Alessandro
  Verri}.} \bibinfo{year}{2018}\natexlab{}.
\newblock \showarticletitle{Latent variable time-varying network inference}. In
  \bibinfo{booktitle}{{\em KDD}}.
\newblock


\bibitem[\protect\citeauthoryear{Wang, Li, Golbandi, Bendersky, and
  Najork}{Wang et~al\mbox{.}}{2018}]%
        {wang2018lambdaloss}
\bibfield{author}{\bibinfo{person}{Xuanhui Wang}, \bibinfo{person}{Cheng Li},
  \bibinfo{person}{Nadav Golbandi}, \bibinfo{person}{Michael Bendersky}, {and}
  \bibinfo{person}{Marc Najork}.} \bibinfo{year}{2018}\natexlab{}.
\newblock \showarticletitle{The lambdaloss framework for ranking metric
  optimization}. In \bibinfo{booktitle}{{\em CIMK}}.
\newblock


\bibitem[\protect\citeauthoryear{Yang, Zhong, Li, and Jie}{Yang
  et~al\mbox{.}}{2017}]%
        {yang2017bi}
\bibfield{author}{\bibinfo{person}{Carl Yang}, \bibinfo{person}{Lin Zhong},
  \bibinfo{person}{Li-Jia Li}, {and} \bibinfo{person}{Luo Jie}.}
  \bibinfo{year}{2017}\natexlab{}.
\newblock \showarticletitle{Bi-directional joint inference for user links and
  attributes on large social graphs}. In \bibinfo{booktitle}{{\em WWW}}.
  \bibinfo{pages}{564--573}.
\newblock


\bibitem[\protect\citeauthoryear{Zhou, Lafferty, and Wasserman}{Zhou
  et~al\mbox{.}}{2008}]%
        {zhou2010time}
\bibfield{author}{\bibinfo{person}{Shuheng Zhou}, \bibinfo{person}{John
  Lafferty}, {and} \bibinfo{person}{Larry Wasserman}.}
  \bibinfo{year}{2008}\natexlab{}.
\newblock \showarticletitle{Time varying undirected graphs}. In
  \bibinfo{booktitle}{{\em COLT}}.
\newblock


\end{thebibliography}
